\documentclass[twocolumn,showpacs,preprintnumbers,amsmath,amssymb]{revtex4}

\usepackage{graphicx}
\usepackage{dcolumn}
\usepackage{bm}

\usepackage[english]{babel}
\usepackage[ansinew]{inputenc}
\usepackage{amsfonts}
\usepackage{latexsym}
\usepackage{amsmath}
\usepackage{amssymb}

\begin{document}

\title{Association of heteronuclear molecules in a harmonic oscillator well}

\author{Jesper Fevre Bertelsen$^1$ and Klaus M\o lmer$^2$}
\affiliation{$^1$Danish National Research Foundation Center for Quantum Optics, Department of Physics and Astronomy, University of
Aarhus, DK-8000 \AA rhus C, Denmark, $^2$Lundbeck Foundation
Theoretical Center for Quantum System Research, Department of
Physics and Astronomy, University of Aarhus, DK-8000 \AA rhus C,
Denmark}

\date{\today}

\begin{abstract}
We describe the collisional interaction between two different atoms that are trapped in a harmonic
potential. The atoms are exposed to a magnetic field, which is modulated in the vicinity of an s-wave
Feshbach resonance, and we study the formation of molecular bound states and excited states of the
trapped system with non-trivial angular correlations.
\end{abstract}


\maketitle

\section{Introduction}

Among the variety of processes and phenomena that have been studied with degenerate quantum gasses,
the conversion of an atomic Bose \cite{Herbig,Xu,RempeMol,Hodby} or Fermi-gas \cite{Hodby,Jin,Jochim}
of atoms into a gas consisting of diatomic molecules is one of the most fascinating, because the
entire system undergoes a dramatic change of state and because it provides a practical way to produce
a quantum degenerate state of a molecular species, which may not be reached easily in any other way.
This process, which can be driven both by photo-association, by sweeping a B-field across a Feshbach
collisional resonance in the system and by RF association from another hyperfine state, has been
studied both in larger trapped samples where the collective many-body state changes character, and in
the Mott insulating phase in optical lattices \cite{Hamburg,Ospelkaus,Innsbruck,Garching,Esslinger}, where the
process involves only two particles and can be understood microscopically.

With the association of different atomic species one has the possibility to form polar molecules,
which offer interesting interaction dynamics both in large condensates
\cite{Santos,Goral,Dell1,Dell2}, and in Mott insulators, where the long range dipole-dipole
interaction between molecules, e.g., makes them an interesting candidate for quantum simulators and
quantum computers \cite{DeMille,Zoller}. In the present theoretical paper we study the dynamics of a pair
of different atoms in the vicinity of an s-wave collisional resonance. We treat the case of atomic
systems prepared in an optical lattice potential with precisely one atom pair per site, and we treat
the case of a deep lattice well approximated by a harmonic potential. This system has recently been
implemented and studied \cite{Hamburg}, and the production of molecules by RF association from another
hyperfine state was demonstrated. We shall identify the eigenstates of the problem for different
values of the B-field and solve the dynamical equations when the field is modulated. In particular we
will show that oscillations of the field with a frequency which is resonant with the Bohr frequency
between discrete atomic-like and molecular-like levels in the trapping potential offers unique
opportunities for controllable transfer of the atoms to specific excited and molecular bound states in
the trap.

In Sec II,we outline the basic theory, which is complicated here by the fact that the problem with
different atoms does not separate in center-of-mass and relative coordinates. In Sec. III, we treat
the atomic interaction by the conventional pseudo-potential, and we show that all necessary matrix
elements of the Hamiltonian can be determined analytically, and the eigenstates and energies can be
found by diagonalization of not too large matrices. The eigenstates have definite total angular
momentum, but the non-separability implies that the states are not eigenstates of the separate
center-of-mass and relative angular momentum. In Sec. IV we generalize our earlier dynamical calculations \cite{Jesper} to the heteronuclear case. Using a harmonic oscillation of the magnetic field we demonstrate association of heteronuclear molecules with 100 \% efficiency and show that even though we start from spherically symmetric states and have only s-wave interaction and symmetric external potentials, it is possible to drive the system into states with non-zero relative angular momentum. Sec. V
concludes the paper.

\section{Non-separable motion of different atoms in a harmonic potential}

If the potential depth in an isotropic 3-dimensional optical lattice, $V(\mathbf
r)=V_0\sum_{i=x,y,z}\sin^2\left(kx_i\right)$, is significantly larger than the tunnelling energy
between lattice sites, atoms placed in the lattice may be prepared in a Mott-insulating state with a
well defined number of atoms in each well. Provided that the lattice is also deeper than the recoil energy
$\hbar^2k^2/2m$, the atoms experience around each potential minimum a harmonic oscillator potential
\begin{align}
V(\mathbf r)\simeq V_0 k^2 \sum_{i=x,y,z} x_i^2=V_0 k^2 r^2= \frac{1}{2}m\omega^2r^2,
\end{align}
where $\omega=\sqrt{2V_0k^2/m}$, $k$ is the optical wave number and $m$ is the mass of
each atom.

We shall treat the dynamics of two atoms, with masses $m_1$ and $m_2$ and position coordinates
$\mathbf r_1$ and $\mathbf r_2$, from which we form the center-of-mass (CM) and relative coordinates
\begin{align}
\mathbf R=\frac{m_1\mathbf r_1+m_2\mathbf r_2}{m_1+m_2}\qquad\mathbf
r=\mathbf r_1-\mathbf r_2.
\end{align}

The Hamiltonian with an atomic interaction potential $V_\textrm{int}(\mathbf r)$ reads
\begin{align}\label{Hamilton}
H=-\frac{\hbar^2}{2m_1}\nabla_1^2-\frac{\hbar^2}{2m_2}\nabla_2^2+
\frac{1}{2}m\omega_1^2r_1^2+\frac{1}{2}m\omega_2^2r_2^2+V_\textrm{int}(\mathbf
r),
\end{align}
and it does not separate in center-of-mass and and relative coordinates, but it can be rewritten
\begin{align}\label{HamiltonianWCoupling1}
H=H_\textrm{cm}(\mathbf R)+H_\textrm{rel}(\mathbf r)+C\mathbf R\cdot\mathbf r,
\end{align}
with
\begin{align}\label{HCM1}
H_\textrm{cm}&=\frac{-\hbar^2}{2M}\nabla_\mathbf R^2+\frac{1}{2}M\Omega^2 R^2\\ \label{Hrel1}
H_\textrm{rel}&=\frac{-\hbar^2}{2\mu}\nabla_\mathbf r^2+\frac{1}{2}\mu\omega^2
r^2+V_\textrm{int}(\textbf{r}).
\end{align}

In (\ref{HCM1}), we have introduced the total and reduced masses $M=m_1+m_2$, $\mu=m_1m_2/M$.

The frequencies $\Omega$ and $\omega$ and the coupling coefficient $C$ have non-trivial values
determined by equating the potential terms in (\ref{Hamilton}) and (\ref{HamiltonianWCoupling1}),
\begin{align}\label{OmegaFormula}
\frac{\Omega}{\omega_1}=\sqrt{\frac{1+\beta}{1+\alpha}},\
\frac{\omega}{\omega_1}=\sqrt{\frac{\alpha+\beta/\alpha}{1+\alpha}},\
\frac{C}{m_1\omega_1^2}=\frac{\alpha-\beta}{1+\alpha}
\end{align}
where we define the mass ratio $\alpha=m_2/m_1$ and the ratio of the harmonic oscillator depths
$\beta=(m_2\omega_2^2)/(m_1\omega_1^2)$.

Note that in the case $\omega_2=\omega_1$ (that is, $\alpha=\beta$) we have
$\Omega=\omega=\omega_1=\omega_2$ and $C=0$ so the center-of-mass and relative motion separate
exactly, independently of the masses. In the general case we obtain a simple $C\mathbf R\cdot\mathbf
r$ coupling of the two degrees of freedom, which suggests to diagonalize this coupling in the factored
basis of eigenstates of the separate problems.

Let us take $^{87}$Rb and $^6$Li as a specific example with a rather large mass ratio. If we take
$\beta=(m_\textrm{Li}\omega_\textrm{Li}^2)/(m_\textrm{Rb}\omega_\textrm{Rb}^2)=0.4$ (which means that
$\omega_\textrm{Li}=2.41\times\omega_\textrm{Rb}$) we have $\Omega=1.14\times\omega_\textrm{Rb}$, $\omega=2.34\times\omega_\textrm{Rb}$, and $C=-0.877\ \mu\omega^2$ from (\ref{OmegaFormula}). Taking the harmonic oscillator length scale as an estimate of
the magnitude of the dipole matrix elements, we can get the following order of magnitude estimate of
the energy shift of the lowest energy states caused by the coupling:
\begin{align}
\frac{|\Delta E|}{\hbar\omega}
\sim\frac{|C|}{\hbar\omega}\sqrt{\frac{\hbar}{M\Omega}}\sqrt{\frac{\hbar}{\mu\omega}}
=\frac{|C|}{\mu\omega^2\sqrt{(M/\mu)(\Omega/\omega)}}=0.3
\end{align}
so the coupling is expected to be significant but the product basis is still a good starting point for
our analysis.

We shall assume that the interaction potential is central, i.e., it is independent of the relative
angular orientation of the atoms. This implies that the  $H_\textrm{cm}$ and $H_\textrm{rel}$  commute
with the angular momentum operators,
\begin{align}
\mathbf L=\mathbf R\times\mathbf P \qquad \mathbf\mathcal l=\mathbf r\times\mathbf p
\end{align}
where $\mathbf P$ and $\mathbf p$ are the conjugated momentum operators of $\mathbf R$ and $\mathbf r$ respectively and we can choose as basis for our description the product states
\begin{align}
\psi_{NLMnlm}(\mathbf R,\mathbf r)=\Phi_{NLM}(\mathbf R)\varphi_{nlm}(\mathbf r)
\end{align}
where $N$ and $n$ are principal quantum numbers and $L,M$ and $l,m$ are the angular momentum quantum
numbers of the CM and relative motion respectively. It is interesting to note that the total angular momentum
of the particles $\mathbf J=\mathbf l_1+\mathbf l_2$ can also be written as
\begin{align}
\mathbf J=\mathbf L+\mathbf\mathcal l.
\end{align}
Since both $H_\textrm{CM}$, $H_\textrm{rel}$ and $\mathbf R\cdot\mathbf r$ are rotationally invariant,
the total angular momentum commutes with the full Hamiltonian, and our Hilbert space separates into
independent subspaces belonging to different values of $J$ and $m_J$ which makes it natural to switch
from the $|NLMnlm\rangle$ basis to the coupled $|Jm_JNLnl\rangle$ basis:
\begin{align}\label{JMStates}
\psi^{Jm_J}_{NLnl}=\sum_{Mm}\langle LMlm|Jm_J\rangle\Phi_{NLM}(\mathbf R)\varphi_{nlm}(\mathbf r)
\end{align}
where $\langle LMlm|Jm_J\rangle$ denote the Clebsch Gordan coefficients.

To diagonalize the $\mathbf R\cdot\mathbf r$ coupling term in the basis (\ref{JMStates}) we need the
matrix elements:
\begin{align}\nonumber
&\langle\psi^{Jm_J}_{N'L'n'l'}|\mathbf R\cdot\mathbf r|\psi^{Jm_J}_{NLnl}\rangle
\\[6pt]\nonumber
&=\sum_{M'm'Mm}\langle L'M'l'm'|Jm_J\rangle\langle LMlm|Jm_J\rangle
\\[6pt]\nonumber
&\quad\times\langle\Phi_{N'L'M'}(\mathbf R)\varphi_{n'l'm'}(\mathbf r)|\mathbf R\cdot\mathbf r|
\Phi_{NLM}(\mathbf R)\varphi_{nlm}(\mathbf r)\rangle
\\[6pt]\label{DipoleExpr1}
\end{align}
It should be emphasized that $\mathbf R\cdot\mathbf r$ does \emph{not} commute with $\mathbf L^2$ and
$\mathbf l^2$ so $L$ and $l$ are not good quantum numbers when the coupling term is included into the
Hamiltonian. This is in contrast to, e.g., the $\mathbf L\cdot\mathbf S$ (spin-orbit) coupling in the
hydrogen atom which commutes with $L^2$ and $S^2$ leaving $L$ and $S$ as good quantum numbers.

\section{Eigenstates of relative motion}

\subsection{Pseudopotential description}

At sufficiently low temperatures atoms interact mainly through s-wave scattering since the centrifugal
barrier for higher partial waves becomes much larger than the thermal energies. The interaction
depends to a large extent only on a single parameter, the s-wave scattering length $a_{sc}$, and the
real physical interaction can be modelled by the following pseudopotential \cite{Huang,Huang&Yang}
\begin{align}\label{pseudopot}
V_\textrm{int}(r)=\frac{2\pi\hbar^2a_\textrm{sc}}{\mu}\delta^{(3)}(\mathbf r)\frac{\partial}{\partial r}r
\end{align}
where the action of the operator $(\partial/\partial r)r$ on a wave function $\psi$ is to be
understood as $(\partial/\partial r)(r\psi)$. The pseudopotential reproduces the correct wave function in the entire range outside the range of the physical interaction.

Writing $u(r)=r\psi(r)$, the pseudopotential (\ref{pseudopot}) implies the boundary condition
\cite{Greene},
\begin{align}\label{bound}
\frac{u'(0)}{u(0)}=\frac{-1}{a_\textrm{sc}}
\end{align}
instead of the usual $u(0)=0$ in a regular potential.

For interatomic potentials containing a long range van der Waals ($V(r)=-C_6/r^6$) attraction, the
validity of the pseudopotential description for trapped atoms in a harmonic oscillator is determined
by the ratio $\beta_6/a_\textrm{ho}$ where $\beta_6=(2\mu C_6/\hbar^2)^{1/4}$ is the characteristic
potential length scale and $a_\textrm{ho}=\sqrt{\hbar/\mu\omega}$ is the length scale of the harmonic
confining potential \cite{BlumePRA65,BoldaPRA66}. Values of $C_6$ for both homonuclear and
heteronuclear alkali metal atom pairs can be found in \cite{Dalgarno}. In the case of an optical
lattice well with a trapping frequency of $\omega_\textrm{Rb}=2\pi\times 200$ kHz we have $a_\textrm{ho}=1.2\times
10^3\ a_0$ and since $\beta_6=85$ a$_0$ for $^{87}$Rb-$^6$Li \cite{Dalgarno} we expect
the pseudopotential approximation to be reasonably good even for a very tightly confining lattice.

The real interaction potential accommodates several bound states, but here we are only concerned with the
very loosely bound state which can be accessed in Feshbach resonance experiments and this bound state
is accounted for by the pseudopotential. The parameter which is used to tune the scattering length in
experiments is the magnitude of the magnetic field $B$. Near a Feshbach resonance the scattering
length is given by \cite{Moerdijk}
\begin{align}\label{Feshbach2}
a_\textrm{sc}=a_\textrm{bg}\left(1-\frac{\Delta}{B-B_0}\right)=a_\textrm{bg}
\left(1-\left(\frac{B-B_0}{\Delta}\right)^{-1}\right)
\end{align}
so if everything is expressed in terms of the dimensionless magnetic field $(B-B_0)/\Delta$ the only
tunable parameters in the pseudopotential model are the harmonic oscillator frequency $\omega$ and the
background scattering length $a_\textrm{bg}$ (notice, however, that $\Delta$ might be negative).

Calculations of spectra in models with more accurate potentials can be found in
\cite{JuliennePRA61,Chen,BlumePRA65}, and pseudopotential models with an energy dependent scattering length have been used to improve the accuracy \cite{BlumePRA65,BoldaPRA66}. While the exact spectra and
matrix elements might be slightly different due to the limitations of our approximation, the
pseudopotential captures the essential physics quite well and it gives good quantitative agreement in
most cases.

\subsection{Wave functions of relative motion}

In this section we obtain solutions to the Schr\"odinger equation of the relative motion with the
Hamiltonian (\ref{Hrel1}) and the pseudopotential (\ref{pseudopot}). These solutions, i.e., the energy
spectrum and the wave functions have been provided previously \cite{Busch,Greene}, but we give
them here for completeness, as we shall use their precise form to subsequently evaluate the matrix
elements of the coupling term $C\mathbf R\cdot\mathbf r$.

Let us first, for reference provide the spectrum
\begin{align}\label{isolevel}
E=\hbar\omega\left(2n+l+\frac{3}{2}\right)
\end{align}
and the associated eigenfunctions
\begin{align}\label{HarmOscWF}
&\psi_{nlm}(\mathbf r)
\\
&=\sqrt{\frac{2n!}{\Gamma(n+l+3/2)}}\ L_{n}^{l+1/2}(r^2)\ r^le^{-r^2/2}\
Y_{lm}(\theta,\phi),
\end{align}
for the three-dimensional isotropic harmonic oscillator. $L_n^\alpha$ are generalized Laguerre
polynomials, $r$ is given in units of the harmonic oscillator length $\sqrt{\hbar/(\mu \omega)}$, and
$Y_{lm}$ are the spherical harmonic wave functions.

The pseudopotential (\ref{pseudopot}) only alters the solutions with $l=0$ \cite{Busch}. In the
region $r>0$ the radial Schr\"odinger equation, expressed in units of the harmonic oscillator length
scale $a_\textrm{ho} =\sqrt{\hbar/\mu\omega}$ and energy scale $E=\hbar\omega\epsilon$, reads
\begin{align}
-\frac{1}{2}\frac{\partial^2 u}{\partial r^2}+\frac{1}{2}r^2 u=\epsilon u
\end{align}
or
\begin{align}
\frac{\partial^2 u}{\partial (\sqrt{2}r)^2}-\left(\frac{1}{4}(\sqrt{2}r)^2-\epsilon\right)u=0
\end{align}
According to \cite{A&S} 19.3.7 and 19.3.8, the solutions to this equation are the parabolic cylinder
functions $D_{\epsilon-1/2}(\sqrt{2}r)$ and $V(-\epsilon,\sqrt{2}r)$ with the following asymptotic behaviour for
large $r$ (\cite{A&S} 19.8.1 and 19.8.2):
\begin{align}
D_{\epsilon-1/2}(\sqrt{2}r)\simeq e^{-r^2/2}(\sqrt{2}r)^{\epsilon-1/2}\\
V(-\epsilon,\sqrt{2}r)\simeq\sqrt\frac{2}{\pi}e^{r^2/2}(\sqrt{2}r)^{-\epsilon-1/2}.
\end{align}
The diverging $V$ function must be discarded. The $D$ function satisfies (\cite{A&S} 19.3.5 and
19.3.7)
\begin{align}\label{DValue0}
D_{\epsilon-1/2}(0)=\frac{\sqrt{\pi}}{2^{-\epsilon/2+1/4}\Gamma(-\epsilon/2+3/4)}
\\[6pt]\label{DDValue0}
D_{\epsilon-1/2}'(0)=\frac{-\sqrt{\pi}}{2^{-\epsilon/2-1/4}\Gamma(-\epsilon/2+1/4)}
\end{align}
Normally we would only accept the regular solutions with $u(0)=0$, which implies that the $\Gamma$
function in (\ref{DValue0}) is infinite, and hence that its argument is a non-positive integer. Thus,
$\epsilon=2n+3/2$ where $n=0,1,2,\ldots$ in agreement with (\ref{isolevel}). The pseudopotential
(\ref{pseudopot}), however, imposes the boundary condition (\ref{bound}), and the discrete energy
spectrum $\epsilon_\nu$ is given by the solutions to the following transcendental equation:
\begin{align}\nonumber\label{TEq1}
\frac{\sqrt{2}D_{\epsilon-1/2}'(0)}{D_{\epsilon-1/2}(0)}
=\frac{-2\Gamma(-\epsilon/2+3/4)}{\Gamma(-\epsilon/2+1/4)}=\frac{-a_\textrm{ho}}{a_\textrm{sc}}
\end{align}
It is convenient to associate to each eigenenergy an effective harmonic oscillator quantum number $\nu$ defined by
\begin{align}
\epsilon=2\nu+\frac{3}{2}
\end{align}
such that $u(r)\propto D_{2\nu+1}(\sqrt{2}r)$ and
\begin{align}\label{HOPPstates1}
\psi_\nu(\mathbf r)=\frac{u_\nu(r)}{r}Y_{00}
=\frac{N_\nu}{r}2^{-\nu-1/2}D_{2\nu+1}(\sqrt{2}r)Y_{00}
\end{align}
where
\begin{align}
N_\nu=\sqrt\frac{2\Gamma(-\nu-1/2)\Gamma(-\nu)}{\pi[\psi(-\nu)-\psi(-\nu-1/2)]}
\end{align}
is a normalization constant and $\psi$ is the digamma function (see Appendix). The wave function generally diverges as $1/r$ for
small $r$ since $u(0)$ is generally finite according to (\ref{DValue0}). The expression (\ref{HOPPstates1}) can also be rewritten in terms of the confluent hypergeometric $U$ function (see \cite{A&S} 13.6.36 and 13.1.29) to obtain the forms cast in \cite{Busch,Greene}.

\subsection{Dipole matrix elements and non-separable eigenstates}\label{SpectrumNS}

In a given $Jm_J$ subspace the matrix representation of the Hamiltonian (\ref{HamiltonianWCoupling1})
in the $|NL\nu l\rangle$ basis, abbreviated by state indices $p,q$, is the sum of diagonal
contributions from $H_\textrm{CM}$ and $H_\textrm{rel}$ and an off-diagonal contribution from the
coupling term $C\mathbf R\cdot\mathbf r$ (see Eq. \ref{DipoleExpr1}):
\begin{align}\nonumber\label{HamiltonMatrix}
H_{pq}&=\hbar\Omega(2N_p+L_p+3/2)\delta_{pq}+\hbar\omega(2\nu_p+l_p+3/2)\delta_{pq}
\\[6pt]\nonumber
&+C\sum_{M_pm_pM_qm_q}\langle L_pM_pl_pm_p|Jm_J\rangle\langle L_qM_ql_qm_q|Jm_J\rangle
\\[6pt]&\quad\times\nonumber
\langle\Phi_{N_pL_pM_p}(\mathbf R)|\mathbf R|\Phi_{N_qL_qM_q}(\mathbf R)\rangle_\mathbf R
\\[6pt]
&\qquad\cdot
\langle\varphi_{\nu_pl_pm_p}(\mathbf r) |\mathbf r|\varphi_{\nu_ql_qm_q}(\mathbf r)\rangle_\mathbf r
\end{align}
where $\langle\ \rangle_\mathbf R$ and $\langle\ \rangle_\mathbf r$ denote integration with respect to $\mathbf R$, respectively $\mathbf r$ and $\delta_{pq}$ is the Kronecker delta.

The dipole integrals separate into radial and angular components. The radial dipole integral for
harmonic oscillator wave functions is given by the following simple formula, see Appendix:
\begin{align}\nonumber
\int_0^\infty & R_{N'(L+1)}(r)\ r\ R_{NL}(r)r^2\textrm{dr}
\\
&=\sqrt{N+L+3/2}\ \delta_{N',N}-\sqrt{N}\ \delta_{N',N-1}
\end{align}
The parabolic cylinder wave functions have $l=0$ so they couple only to the $l=1$ harmonic oscillator
states. In the Appendix we denote the radial parts of the wave functions by $R_\nu$ and $R_{n1}$, and we
derive
\begin{align}\nonumber
\int_0^\infty & R_\nu(r)r R_{n1}(r) r^2 \mathrm{dr}
\\[6pt]\nonumber
&=N_\nu\ \sqrt\frac{2n!}{\Gamma(n+5/2)}\ \frac{\sqrt\pi}{16}
\\[6pt]
&\ \times\sum_{k=0}^{n} 2^{-2k}
(-1)^k\left(\begin{array}{l}n+\alpha\\n-k\end{array}\right)\frac{1}{k!} \
\frac{\Gamma(2k+4)}{\Gamma(k+2-\nu)}.
\end{align}

We now have analytical expressions for the full Hamiltonian matrix. Consider the case where the system
is prepared, and therefore remains, in the block of $J=0,m_J=0$ states. Such states involve CM and
relative states with $L=l$ and $M=-m$ and using the Clebsch-Gordan coefficient\newline $\langle
LMlm|J=0\ m_J=0\rangle=\delta_{M(-m)}\ (-1)^{L-M}/\sqrt{2L+1}$ we have
\begin{align}\label{J0m0state}\nonumber
&\psi^{00}_{NLnL}(\mathbf R,\mathbf r)
\\
&=\frac{1}{\sqrt{2L+1}}\
\sum_{M=-L}^L(-1)^{L-M}\Phi_{NLM}(\mathbf R)\varphi_{nL(-M)}(\mathbf r).
\end{align}

\begin{figure}[!htb]
\begin{center}
\includegraphics[width=\columnwidth]{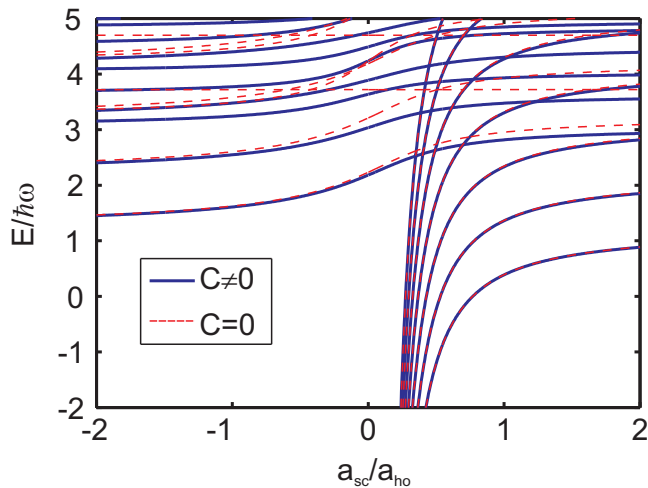}
\caption{\small (Color online) $J=0$ spectrum of a Rb and a Li atom in a harmonic well in the pseudopotential
approximation. The energy scale $\hbar\omega$ of the relative motion harmonic oscillator is chosen as
the unit of energy. $\beta=(m_\textrm{Li}\omega_\textrm{Li}^2)/(m_\textrm{Rb}\omega_\textrm{Rb}^2)=0.4$. The spectrum of
the full Hamiltonian including the coupling between CM and relative dynamics is shown as the
connected, blue lines. The spectrum we get before including the coupling term is shown as red, dashed
lines. We have included states up to $N=6$, $n=6$ and $L=l=12$ which gives a total of $637$ product
basis states. Note the equidistant CM vibrational states of molecular nature when $a_\textrm{sc}>0$.}
\label{RbLiSpectrum1}
\end{center}
\end{figure}

When we diagonalize the Hamiltonian in the basis (\ref{J0m0state}) we get the $J=0$ spectrum
shown in Fig. \ref{RbLiSpectrum1}. Note that we plot the energies as functions of scattering length,
which in turn must be described as a function of the magnetic field in an experiment.

As the eigenfunctions $\Psi$ of the complete Hamiltonian are expanded on the basis (\ref{J0m0state}) of angular momentum
eigenstates
\begin{align}\label{expansion1}
\Psi=\sum_{NLn} c_{NLn}\psi^{00}_{NLnL}
\end{align}
the expansion coefficients provide, through $|c_{NLn}|^2$, the CM and relative angular momentum content of the
eigenstates. We observe that with a central interaction potential and spherically symmetric single-particle confining potentials, it is possible to generate states with non-vanishing relative, and
center-of-mass, angular momenta.  In Tab. \ref{WFcontentTab0} and in Tab. \ref{WFcontentTab0p36} the
population of different angular momentum components are given for the lowest eigenstates of the system for
$a_\textrm{sc}=0$ and for $a_\textrm{sc}/a_\textrm{ho}=0.36$. For instance the state with an energy of
$3.64\ \hbar\omega$ at $a_\textrm{sc}=0$ contains $63.2$ \% of $|010\rangle$ with $L=l=1$, $13.6$ \%
of $|100\rangle$ with $L=l=0$, and $23.1$ \% of other product states.

\begin{table}[!htb]
\begin{tabular}{r@{.}l|lr@{.}l|lr@{.}l}
\multicolumn{2}{c|}{Energy} & \multicolumn{3}{c|}{Largest content} & \multicolumn{3}{c}{Second largest content}
\\
\multicolumn{2}{c|}{} & $|NLn\rangle$ & \multicolumn{2}{c|}{$|c_{NLn}|^2\ (\%)$} & $|NLn\rangle$ & \multicolumn{2}{c}{$|c_{NLn}|^2\ (\%)$}
\\\hline
2&18 & $|000\rangle$ & 96&2 & $|010\rangle$ & 3&5
\\\hline
3&04 & $|100\rangle$ & 79&1 & $|010\rangle$ & 11&2
\\\hline
3&64 & $|010\rangle$ & 63&2 & $|100\rangle$ & 13&6
\\\hline
3&89 & $|200\rangle$ & 64&0 & $|110\rangle$ & 16&7
\\\hline
4&24 & $|001\rangle$ & 80&8 & $|010\rangle$ & 12&6
\\\hline
4&49 & $|110\rangle$ & 33&5 & $|200\rangle$ & 22&8
\\\hline
4&74 & $|300\rangle$ & 51&1 & $|210\rangle$ & 18&3
\end{tabular}
\caption{\small Angular momentum basis content at $a_\textrm{sc}=0$ of the eigenstates of the full
Hamiltonian for $^{87}$Rb and $^{6}$Li in a harmonic potential with
$\beta=(m_\textrm{Li}\omega_{Li}^2)/(m_\textrm{Rb}\omega_{Rb}^2)=0.4$.} \label{WFcontentTab0}
\end{table}

\begin{table}[!htb]
\begin{tabular}{r@{.}l|lr@{.}l|lr@{.}l}
\multicolumn{2}{c|}{Energy} & \multicolumn{3}{c|}{Largest content} & \multicolumn{3}{c}{Second largest content}
\\
\multicolumn{2}{c|}{} & $|NLn\rangle$ & \multicolumn{2}{c|}{$|c_{NLn}|^2\ (\%)$} & $|NLn\rangle$ & \multicolumn{2}{c}{$|c_{NLn}|^2\ (\%)$}
\\\hline
-3&09 & $|000\rangle$ & 99&997 & $|010\rangle$ & 0&0013
\\\hline
-2&12 & $|100\rangle$ & 99&994 & $|110\rangle$ & 0&002
\\\hline
-1&14 & $|200\rangle$ & 99&990 & $|210\rangle$ & 0&003
\\\hline
-0&16 & $|300\rangle$ & 99&985 & $|310\rangle$ & 0&004
\\\hline
0&81 & $|400\rangle$ & 99&981 & $|410\rangle$ & 0&005
\\\hline
1&79 & $|500\rangle$ & 99&977 & $|410\rangle$ & 0&006
\\\hline
2&53 & $|001\rangle$ & 91&0 & $|010\rangle$ & 8&0
\\\hline
2&76 & $|600\rangle$ & 99&973 & $|510\rangle$ & 0&008
\\\hline
3&29 & $|101\rangle$ & 58&7 & $|010\rangle$ & 22&0
\\\hline
3&81 & $|010\rangle$ & 58&9 & $|101\rangle$ & 21&7
\\\hline
4&11 & $|201\rangle$ & 43&8 & $|110\rangle$ & 18&8
\\\hline
4&57 & $|002\rangle$ & 64&0 & $|011\rangle$ & 17&1
\\\hline
4&71 & $|110\rangle$ & 26&4 & $|201\rangle$ & 16&9
\\\hline
4&93 & $|301\rangle$ & 30&6 & $|201\rangle$ & 17&2
\end{tabular}
\caption{\small Angular momentum basis content at $a_\textrm{sc}=0.36\sqrt{\hbar/\mu\omega}$ of the
eigenstates of the full Hamiltonian for $^{87}$Rb and $^{6}$Li in a harmonic potential with
$\beta=(m_\textrm{Li}\omega_{Li}^2)/(m_\textrm{Rb}\omega_{Rb}^2)=0.4$. The molecular states are almost
pure product states with $n=0$, $L=0$ and a well defined CM excitation $N$ whereas for all other
states the CM dynamics and the relative dynamics are entangled.} \label{WFcontentTab0p36}
\end{table}

\section{Resonant Dynamics near a Feshbach resonance}\label{DynamicsChap}

A recent experiment \cite{Ospelkaus} has demonstrated the formation of heteronuclear molecules by RF association from another hyperfine state near a Feshbach resonance in the $^{87}$Rb-$^{40}$K system.

Extending our previous work \cite{Jesper} for identical systems we will now address the dynamics beyond the
adiabatic approximation, where a modulation of the field causes transitions \emph{between} the
adiabatic eigenstates. As a specific application we will discuss resonant transitions driven by an
oscillating magnetic field. The idea, proposed and used by Thompson et al. \cite{Thompson} and Greiner
et al. \cite{GreinerJin} and theoretically analyzed in \cite{Jesper,Hanna}, is to apply small sinusoidal oscillations of the magnetic field around a
fixed value $B'$ close to a Feshbach resonance:
\begin{align}\label{BOsc1}
B(t)=B'+b\sin(\omega_B t).
\end{align}
If $\hbar\omega_B$ matches the energy difference between two stationary states,  it is possible to
transfer population from one state to another (for instance to the molecular state) or to create well
controlled quantum mechanical superposition states.

The modulation technique has also been used to produce and make spectroscopy on ultracold
heteronuclear $^{87}$Rb-$^{85}$Rb molecules \cite{Papp}, to probe the excitation spectrum of a Fermi
gas in the BEC-BCS crossover regime \cite{GreinerJin}, and to infer the lifetime of short-lived
Feshbach molecules by looking at the width of the resonance \cite{Gaebler}. It differs in principle
from the traditional RF spectroscopy used in e.g. \cite{Hamburg,Regal,Gupta} in that the atoms are
not transferred to other hyperfine levels and in that it is the tunable interatomic interaction that
drives the transition.

When the magnetic field is varied in time, the interaction Hamiltonian is explicitly time dependent,
and we have different options to determine the evolution of the system.

\subsection{Weak  amplitude modulations of B-field}\label{WeakAmplSec}

If the B-field modulation is weak, it constitutes only a small time-dependent perturbation on the
system, and a calculation naturally departs from an expansion in the adiabatic eigenstate basis. We
label the complete set of eigenstates of the Hamiltonian at a given field strength by
$|\Psi_q(B)\rangle$ and their energies $E_q$, and we expand the wave function $|\Psi(t)\rangle$
\begin{align}\label{expansion}
|\Psi(t)\rangle=\sum_q c_q(t)|\Psi_q(B(t))\rangle
\end{align}
Inserting into the time-dependent Schr\"odinger equation, and projecting onto the orthogonal states
$\langle\Psi_p(B(t))|$ we get the coupled amplitude equations
\begin{align}\label{DynEq2}\nonumber
i\hbar\frac{d c_p(t)}{d t} & =E_p(t) c_p(t)
\\&
-i\hbar\sum_q \left.\frac{\partial}{\partial
t'}\langle\Psi_p(B(t))|\Psi_q(B(t'))\rangle\right|_{t'=t}\ c_q(t).
\end{align}
If we take the sinusoidal magnetic field dependence (\ref{BOsc1}) in (\ref{DynEq2}) we get the
following equation for the coefficients
\begin{align}\label{DynEq4}
i\hbar\frac{d c_p(t)}{d t}=E_p(t) c_p(t)-i\hbar\cos(\omega_B t)\sum_q\Omega_{pq}(B(t))\ c_q(t).
\end{align}
where the Rabi frequencies $\Omega_{pq}$ are given by
\begin{align}
\Omega_{pq}(B)=b\omega_B\left.\frac{\partial}{\partial
B'}\langle\Psi_p(B)|\Psi_q(B')\rangle\right|_{B'=B}.
\end{align}

For a weakly modulated field, $b\ll\Delta$, the time dependence of $E_p$ and $\Omega_{pq}$ can be
neglected and the modulation is equivalent to, e.g., the radiative coupling of atomic energy levels.
If $\hbar\omega_B$ is close to the energy difference between two eigenstates $p$ and $q$, and we have
all the population in the states $p$ initially, we therefore expect to see Rabi oscillations between
the two states
\begin{align}\label{RabiOsc1}
|c_q(t)|^2&=1-|c_p(t)|^2=\left(\frac{\Omega_{pq}}{\Omega_{pq}'}\right)^2\sin^2\left(\frac{\Omega_{pq}'\
t}{2}\right)
\end{align}
where
\begin{align}
\Omega_{pq}'=\sqrt{\Omega_{pq}^2+(\omega_B-(E_p-E_q)/\hbar)^2}.
\end{align}

\subsection{General dynamical equations in modulated B-field}\label{DynEqSec}

The expansion  (\ref{expansion}) is general and may be applied for any modulation of the magnetic
field, but since the eigenstates in that expansion have to be found first by numerical diagonalization
of the coupling Hamiltonian, we find it more convenient to use instead the basis (\ref{JMStates}),
where only the relative coordinate part of the wave function depends explicitly on time through the
scattering length. We then get
\begin{align}\label{DynEq3}
&i\hbar\frac{d c_p(t)}{d t}
\\[6pt]
&=\sum_q H_{pq}(t)c_q(t)- i\hbar\sum_q \left.\frac{\partial}{\partial
t'}\langle\psi_p(t)|\psi_q(t')\rangle\right|_{t'=t}\ c_q(t)
\end{align}
where $H_{pq}$ is given in (\ref{HamiltonMatrix}).

The explicit expression for the time derivative of $\langle\psi_p(t)|\psi_q(t')\rangle$ in the basis
(\ref{JMStates}) is
\begin{align}\nonumber\label{TimeDeriv1}
&\left.\frac{\partial}{\partial t'}\langle\psi_p(t)|\psi_q(t')\rangle\right|_{t'=t}
\\[6pt]\nonumber
&=\sum_{M_p m_p M_q
m_q}\langle L_pM_pl_pm_p|J m_J\rangle\langle L_qM_ql_qm_q|J m_J\rangle
\\[6pt]\nonumber
&\quad\times
\langle\Phi_{N_p L_p M_p}|\Phi_{N_q L_q M_q}\rangle_\mathbf R
\\[6pt]
&\quad\times
\left.\frac{\partial}{\partial t'}\langle\varphi_{n_p l_p m_p}(t)|\varphi_{n_q l_q m_q}(t')\rangle_\mathbf r\right|_{t'=t}
\end{align}
where we have exploited the fact that only the relative coordinate wave function $\varphi$ is time
dependent. We note that our use of the adiabatic eigenfunctions of $H_\textrm{rel}$ removes the need to
numerically compute any matrix elements of the singular interaction potential, and thus we make
optimum use of our knowledge of the analytical, irregular solutions to the pseudopotential equation.

It is a great advantage of the pseudopotential model that the matrix element $\langle\varphi_{\nu
00}|\varphi_{\nu' 00}\rangle$ can be calculated analytically, and we can hence express the time
derivative using the chain rule:
\begin{align}\nonumber\label{ChainRule1}
&\left.\frac{\partial}{\partial t'}\langle\varphi_{n_p 0 0}(t)|\varphi_{n_q 0 0}(t')\rangle_\mathbf
r\right|_{t'=t}
\\[6pt]
&=\frac{dB}{dt}\ \frac{da_\textrm{sc}}{dB}\
\left(\frac{da_\textrm{sc}}{d\nu'}\right)^{-1}
\left.\frac{\partial}{\partial\nu'}\langle\varphi_{\nu_p 0 0}|\varphi_{\nu' 0 0}\rangle_\mathbf
r\right|_{\nu'=\nu_q}
\end{align}
where $da_\textrm{sc}/dB$ is given by
\begin{align}\label{ChainRule2}
\frac{da_\textrm{sc}}{dB}=\frac{-\Delta}{(B-B_0)^2}\ a_\textrm{bg},
\end{align}
and $da_\textrm{sc}/d\nu$ is given by
\begin{align}\label{ChainRule3}
\frac{da_\textrm{sc}}{d\nu} =\frac{a_\textrm{sc}}{[f(\nu)]^2}=\frac{a_\textrm{bg}}{[f(\nu)]^2}\
\frac{B-B_0-\Delta}{B-B_0}
\end{align}
with
\begin{align}
f(\nu)=\frac{1}{\sqrt{\psi(-\nu)-\psi(-\nu-\frac{1}{2})}}.
\end{align}
The overlap $\langle\varphi_{\nu 00}|\varphi_{\nu' 00}\rangle$ is provided in the Appendix and all
coefficients in our coupled set of equations are thus given by analytical expressions.

\subsection{Results}\label{AssHetMolSec}
According to the discussion in Sec. \ref{WeakAmplSec} a small amplitude harmonic oscillation of the magnetic field at a frequency such that $\hbar\omega$ is resonant with the energy difference between two eigenstates will inevitably lead to full contrast Rabi oscillations as long as there is a finite coupling between the two states. For $\mathbf J=0$ such a finite coupling arises whenever the basis expansion (\ref{expansion1}) of the two eigenstates at the given magnetic field $B'$ contains basis functions of the form $\Phi_{N 0 0}(\mathbf R)\varphi_{\nu 0 0}(\mathbf r)$ respectively $\Phi_{N 0 0}(\mathbf R)\varphi_{\nu' 0 0}(\mathbf r)$, see Eq. \ref{TimeDeriv1}.

\begin{figure}[!htb]
\begin{center}
\includegraphics[width=\columnwidth]{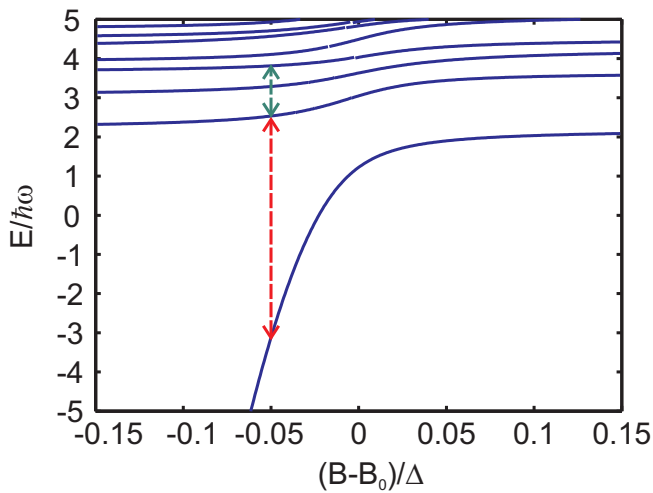}
\caption{\small (Color online) Spectrum for $^{87}$Rb-$^6$Li in an optical lattice well with $\omega_\textrm{Rb}=2\pi\times 200$ kHz and $\beta=m_\textrm{Li}\omega_\textrm{Li}^2/(m_\textrm{Rb}\omega_\textrm{Rb}^2)=0.4$ in the vicinity of a Feshbach resonance with $a_\textrm{bg}=20\ a_0$. Molecular states that have a CM excitation ($N>0$) are not shown. Association of a heteronuclear molecule is indicated with the lower, red arrow while transfer to a state with angular momentum excitation is illustrated with the upper, green arrow.}
\label{E_B_RbLi_Fig}
\end{center}
\end{figure}

\begin{figure}[!htb]
\begin{center}
\includegraphics[width=\columnwidth]{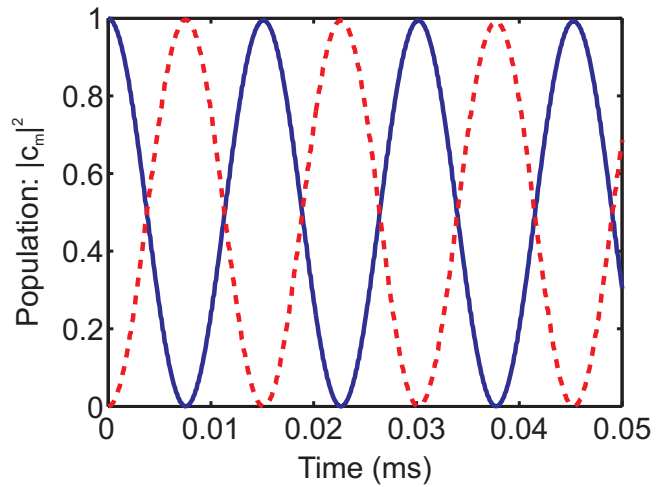}
\caption{\small (Color online) Resonant association of heteronuclear molecules by an oscillating magnetic field. The time dependent magnetic field is $B(t)=B_0-0.05\Delta+0.005\Delta\sin(\omega_B t)$ where $\omega_B=2\pi\times 2660$ kHz. The blue line shows the population of the atomic ground state whereas the population of the molecular ground state is shown with the red, dashed line. $\omega_\textrm{Rb}=2\pi\times 200$ kHz, $\beta=m_\textrm{Li}\omega_\textrm{Li}^2/(m_\textrm{Rb}\omega_\textrm{Rb}^2)=0.4$. We have included states up to $N=3$, $n=3$ and $L=l=6$.}
\label{MolAssociationFig}
\end{center}
\end{figure}

This means e.g. that efficient association of heteronuclear molecules should be possible. We can illustrate it in principle using the $^{87}$Rb-$^6$Li system treated above. The scattering length at zero magnetic field was measured by C. Silber et al. \cite{Silber} to be $20(+9/-6)\ a_0$. A Feshbach resonance with this background scattering length gives rise to the energy spectrum shown in Fig. \ref{E_B_RbLi_Fig}. If we oscillate the magnetic field resonantly around $B=B_0-0.05\Delta$ (corresponding to $a_\textrm{sc}/a_\textrm{ho}=0.36$) we can produce full contrast Rabi oscillations between the lowest non-molecular state and the molecular ground state (illustrated with the lowermost arrow in the spectrum). The dynamics is shown in Fig. \ref{MolAssociationFig}. In practise one might want to use an adiabatic passage to make a robust full transfer as demonstrated in \cite{Jesper}.

Another interesting fact is that the coupling of the CM and relative motion makes it possible to populate states with nonzero CM and relative angular momentum by oscillating the magnetic field close to a Feshbach resonance. It is intriguing that in this way a purely central interatomic interaction can be used to put angular momentum into the relative motion at the cost of exciting the CM motion as well.

\begin{figure}[!htb]
\begin{center}
\includegraphics[width=\columnwidth]{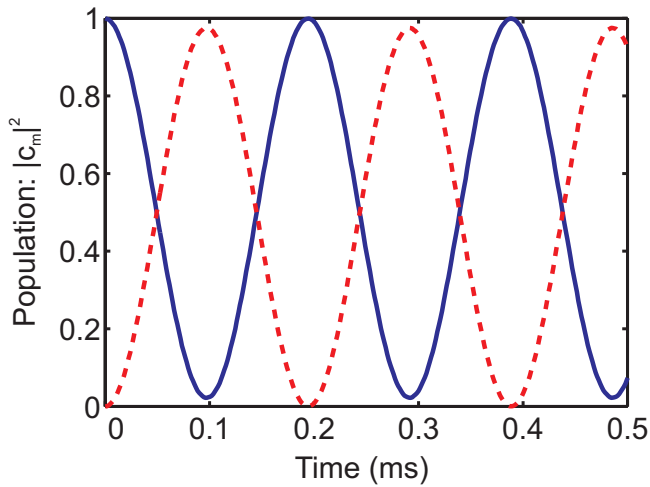}
\caption{\small (Color online) Rabi oscillations between the atomic ground state and a state with $59$ \% content of $L=l=1$. The time dependent magnetic field is $B(t)=B_0-0.05\Delta+0.005\Delta\sin(\omega_B t)$ where $\omega_B=2\pi\times 601$ kHz. The blue line shows the population of the atomic ground state whereas the population of the excited angular momentum state is shown with the red, dashed line. The other parameters are as in Fig. \ref{MolAssociationFig}.}
\label{L1AssociationFig}
\end{center}
\end{figure}

For example the uppermost arrow in Fig. \ref{E_B_RbLi_Fig} illustrates a transfer from the non-molecular atomic ground state to a state with $59$ \% content of $L=l=1$ (see Tab. \ref{WFcontentTab0p36}). If the magnetic field is oscillated resonantly, Rabi oscillations between these two states can be produced (Fig. \ref{L1AssociationFig}) and again a $\pi$-pulse or an adiabatic passage could in principle be used to make a full transfer.

\section{Discussion}

In summary, we have found the eigenstates of motion of a pair of harmonically trapped, different atoms
interacting via a short range central potential. We have investigated the dynamics of the system when
the interaction strength is modulated, by changing the strength of an applied magnetic field, and in
particular we have found that resonant transitions between the molecular and atomic states can be
coherently driven in this system. Interestingly we find that  even though the interaction is central,
states with non-vanishing CM and relative angular momentum of the atoms can be produced. Although our
calculations were done explicitly for the pseudo-potential (\ref{pseudopot}), this is not a formal
restriction on the separation of the problem and the methods applied, and other central interaction
potentials are readily treated by a similar analysis.

We believe that the transition dynamics analyzed in the present paper may be studied with current
experimental techniques. Our work identifies a way to produce molecules without crossing the Feshbach
resonance, and we imagine that the states with nonvanishing relative angular momentum may lead to interesting momentum distributions \cite{Mark,Rempe2,Rempe3} of the individual species when the atoms are released from the lattice.

\acknowledgments{} We gratefully acknowledge discussions with Michael Budde, Nicolai Nygaard and
Ejvind Bonderup.

\section{Appendix: Calculation of matrix elements}\label{CalcMatEl}

This appendix derives the matrix elements used in the Feshbach molecule problem. We consider the
radial wave functions for an isotropic 3-dimensional harmonic oscillator (\ref{HarmOscWF})
\begin{align}
R_{nl}(r)=N_{nl}L_{n}^{l+1/2}(r^2)\ r^le^{-r^2/2}
\end{align}
where $L$ is a generalized Laguerre polynomial, and the s-wave solutions to the problem of a particle
in a harmonic oscillator modified by a regularized s-wave $\delta$-function potential at the origin
(\ref{HOPPstates1})
\begin{align}\label{ParCylWF}
R_\nu (r)=\frac{N_\nu}{r} 2^{-\nu-1/2}D_{2\nu+1}(z),\qquad z=\sqrt{2}r
\end{align}
where $D$ is the parabolic cylinder function. $N_{nl}$ and $N_\nu$ are normalization constants.

\subsection{Radial integral for harmonic oscillator wave functions}

The normalization constant of the harmonic oscillator wave functions is given by
\begin{align}\nonumber
\frac{1}{N_{nl}^2}&=\int_0^\infty\left[L_{n}^{l+1/2}(r^2)\right]^2 r^{2l+2}e^{-r^2}\mathrm{dr}
\\[6pt]\nonumber
&=\frac{1}{2}\int_0^\infty\left[L_{n}^{l+1/2}(u)\right]^2 u^{l+1/2}e^{-u}\mathrm{du}
\\[6pt]\label{RadHarmOscNorm}
&=\frac{\Gamma(n+l+3/2)}{2n!}
\end{align}
where we have used (\cite{A&S} 22.2.12, \cite{G&R} 7.414.3):
\begin{align}\nonumber
\int_0^\infty & e^{-x}x^\alpha L_n^\alpha(x)L_m^\alpha(x)\mathrm{dx}
\\[6pt]
&=\left\{\begin{array}{l}
\frac{\Gamma(\alpha+n+1)}{n!},\quad m=n, \mathrm{Re }\ \alpha > 0
\\[6pt]\label{LaguerreOrth}
0,\qquad\qquad\ \,m\neq n, \mathrm{Re }\ \alpha>-1
\end{array}
\right.
\end{align}
\\

The radial part of the dipole matrix element between oscillator states involves the integral
\begin{align}\nonumber
&\int_0^\infty L_{n'}^{l'+1/2}(r^2)\ L_{n}^{l+1/2}(r^2)\ r^{l'+l+3}e^{-r^2}\mathrm{dr}
\\[6pt]
=\frac{1}{2}&\int_0^\infty L_{n'}^{l'+1/2}(u)\ L_{n}^{l+1/2}(u)\ u^{(l'+l+2)/2}e^{-u}\mathrm{du}\qquad
\end{align}
Since the angular integral has the selection rule $|\Delta l|=1$ we can, without loss of generality,
assume that $l'=l+1$ and employ the recursion relation
\begin{align}
L_n^{\alpha-1}(u)=L_n^{\alpha}(u)-L_{n-1}^{\alpha}(u)
\end{align}
for the generalized Laguerre polynomials (\cite{A&S} 22.7.30) to obtain
\begin{align}\nonumber
&\frac{1}{2}\int_0^\infty L_{n'}^{l'+1/2}(u)\ L_{n}^{l+1/2}(u)\ u^{(l'+l+2)/2}e^{-u}\mathrm{du}
\\[6pt]\nonumber
=&\frac{1}{2}\int_0^\infty L_{n'}^{l+3/2}(u)\ \left(L_n^{l+3/2}(u)-L_{n-1}^{l+3/2}(u)\right)
\\[6pt]\nonumber
&\quad\times u^{(2l+3)/2}e^{-u}\mathrm{du}
\\[6pt]
=&\frac{1}{2}\frac{\Gamma(n'+l+5/2)}{(n')!}\left(\delta_{n',n}-\delta_{n',n-1}\right)
\end{align}
where (\ref{LaguerreOrth}) was used in the last step. Including the normalization
(\ref{RadHarmOscNorm}) we obtain
\begin{align}\nonumber
\int_0^\infty & R_{n',l+1}(r)rR_{nl}(r)r^2\textrm{dr}
\\[6pt]\nonumber
&=\sqrt{\frac{\Gamma(n'+l+5/2)}{\Gamma(n+l+3/2)}\frac{n!}{(n')!}}\ (\delta_{n',n}-\delta_{n',n-1})
\\[6pt]\label{HORadialInt}
&=\sqrt{n+l+3/2}\ \delta_{n',n}-\sqrt{n}\ \delta_{n',n-1}
\end{align}
where the identity $\Gamma(x+1)=x\ \Gamma(x)$ has been used.

\subsection{Radial overlap of parabolic cylinder functions}

The normalization constant of the parabolic cylinder wave functions (\ref{ParCylWF}) is given by
\begin{align}\nonumber
\frac{1}{N_\nu^2}&=\int_0^\infty R_\nu(r)^2 r^2 \mathrm{dr} =2^{-2\nu-1}\int_0^\infty
\left[D_{2\nu+1}(z)\right]^2\frac{\mathrm{dz}}{\sqrt{2}}
\\\nonumber&
=2^{-2\nu-3}\sqrt{\pi}\ \ \frac{\psi(-\nu)-\psi(-\nu-1/2)}{\Gamma(-2\nu-1)}
\\&\label{ParCylNormExpr2}
=\frac{\pi}{2}\ \ \frac{\psi(-\nu)-\psi(-\nu-1/2)}{\Gamma(-\nu-1/2)\Gamma(-\nu)}
\end{align}
where \cite{G&R} 7.711 has been used to solve the integral and the duplication formula for gamma
functions \cite{A&S} 6.1.18 has been used in the last equality. $\psi$ is the digamma function:
\begin{align}\label{DigammaDef}
\psi(x)=\frac{d}{dx}\ln\Gamma(x)=\frac{\Gamma'(x)}{\Gamma(x)}.
\end{align}

In the case $\nu'\neq\nu$ we apply \cite{G&R} 7.711 to determine the overlap integral
\begin{widetext}
\begin{align}\nonumber
\langle\varphi_{\nu'}|\varphi_\nu\rangle
&=N_{\nu'}N_\nu 2^{-\nu'-\nu-1}\int_0^\infty
D_{2\nu'+1}(z)D_{2\nu+1}(z)\frac{\textrm{dz}}{\sqrt 2}
\\[6pt]\nonumber
&=N_{\nu'}N_\nu\frac{\pi}{2(\nu'-\nu)}\left[\frac{1}{\Gamma(-\nu')\Gamma(-\nu-1/2)}-\frac{1}{\Gamma(-\nu)\Gamma(-\nu'-1/2)}\right]
\\[6pt]
&=\frac{f(\nu')f(\nu)}{\nu-\nu'}\times\frac{\Gamma(-\nu')\Gamma(-\nu-\frac{1}{2})-\Gamma(-\nu)\Gamma(-\nu'-\frac{1}{2})}{\sqrt{\Gamma(-\nu'-\frac{1}{2})\Gamma(-\nu')}\
\sqrt{\Gamma(-\nu-\frac{1}{2})\Gamma(-\nu)}}
\end{align}
\end{widetext}
where
\begin{align}\label{ParCylInt2}
f(\nu)=\frac{1}{\sqrt{\psi(-\nu)-\psi(-\nu-\frac{1}{2})}}
\end{align}

\subsection{Dipole matrix element between parabolic cylinder and harmonic oscillator p-wave
functions}\label{PCHOIntSec}

We evaluate
\begin{widetext}
\begin{align}\nonumber
&\int_0^\infty R_\nu(r)R_{n1}(r) r^3\mathrm{dr}
\\[6pt]\nonumber
=&\int_0^\infty \left[\frac{N_\nu}{r} 2^{-\nu-1/2}D_{2\nu+1}(z)\right]\left[N_{n1}L_{n}^{3/2}(r^2)\ r\
e^{-r^2/2}\right] r^3 \mathrm{dr},\qquad z^2=2r^2
\\[6pt]
=&N_\nu N_{n1} 2^{-\nu-1/2}\int_0^\infty D_{2\nu+1}(z)\ L_{n}^{3/2}(z^2/2)\ \frac{z^3}{2^{3/2}}\
e^{-z^2/4}\frac{\mathrm{dz}}{\sqrt 2}
\end{align}
\end{widetext}
The generalized Laguerre polynomials are given by (\cite{A&S} 22.3.9)
\begin{align}\label{LaguerreCoeff}
L_n^\alpha(u)=\sum_{k=0}^n a_k u^k
,\qquad a_k=(-1)^k\left(\begin{array}{l}n+\alpha\\n-k\end{array}\right)\frac{1}{k!}
\end{align}
and we integrate term by term using  (\cite{G&R} 7.722 for $\mu > 0$):
\begin{align}
\int_0^\infty e^{-x^2/4}x^{\mu-1}D_{-\tilde{\nu}}(x)\mathrm{dx}
=\frac{\sqrt\pi 2^{-\mu/2-\tilde{\nu}/2}\Gamma(\mu)}{\Gamma(\mu/2+\tilde{\nu}/2+1/2)}
\end{align}
with $\tilde\nu=-(2\nu+1)$ to get
\begin{align}\nonumber
&\int_0^\infty R_\nu(r)R_{n1}(r) r^3\mathrm{dr}
\\[6pt]
&=N_\nu N_{n1}\frac{\sqrt\pi}{16}\sum_{k=0}^{n} 2^{-2k}
(-1)^k\left(\begin{array}{l}n+\alpha\\n-k\end{array}\right)\frac{1}{k!} \
\frac{\Gamma(2k+4)}{\Gamma(k+2-\nu)}.
\end{align}

\bibliography{refs}

\end{document}